\documentclass[preprint]{aastex6} %delete "trackchanges" for submission 20160519
\usepackage{natbib}
\usepackage[version=3]{mhchem}
\turnoffedit
\shorttitle{ASTE Observation of Jupiter's Stratospheric Composition: Detection of Carbon Monosulfide ($J$=7--6) in 19 Years After the Cometary Impact}
\shortauthors{Iino et al.}

\begin{document}


\title{\edit1{Sub-millimeter Observation of Jupiter's Stratospheric Carbon Monosulfide ($J$=7--6) 19 Years after the Cometary Impact}}


\author{T. Iino\altaffilmark{1}}
\affil{Nature and Science Museum, Tokyo University of Agriculture and Technology (TUAT), Tokyo Japan}
\email{iino@nagoya-u.jp}

\author{H. Ohyama}
\affil{Institute for Space-Earth Environment Research (ISEE), Nagoya University, Aichi Japan}

\author{Y. Hirahara}
\affil{Graduate School of Environment, Nagoya University, Aichi Japan}

\author{T. Takahashi}
\affil{The University of Electro-Communications, Tokyo Japan}

\and

\author{T. Tsukagoshi \altaffilmark{2}}
\affil{School of Science, Ibaraki University, Ibaraki Japan}

%% are identified by the \altaffilmark after each name. Specify alternate

\altaffiltext{1}{Terahertz Research Center, National Institute of Information and Communications Technology, Tokyo Japan }
\altaffiltext{2}{National Institute of Polar Research, Tokyo Japan, Tokyo Japan }


\begin{abstract}
\replaced{On}{In} Jupiter's stratosphere, gaseous carbon monosulfide (CS)\deleted{molecule} was first discovered in 1994 by millimeter and ultraviolet observations as a product induced by the collision of comet Shoemaker-Levy 9 (SL9\deleted{ event}). \replaced{For the new constraint of the}{To constrain} sulfur chemistry, in 2013, 19 years after the SL9 event, we observed Jupiter's stratospheric CS $J$=7 -- 6 rotational transition at 0.8 mm wavelength by using the Atacama Submillimeter Telescope Experiment (ASTE) 10-m single dish telescope. The CS molecular line was successfully detected with 120 mK intensity in the antenna temperature scale. \deleted{Since the lifetime of CS photolysis is as short as 50-100 days in Jupiter's orbit, presence of CS suggests that CS is being recycled efficiently in Jupiter's stratospheric environment.} The \edit1{obtained} CS total mass shows $\sim$90$\%$ decrease relative to that observed in 1998. From the line shape analysis, CS is suggested to be present above the 0.2$^{+0.4}_{-0.15}$ mbar pressure level, which is comparable to that of determined in 1998. \deleted{\ce{H2O}, \replaced{which is considered to be produced by the last cometary impact}{which was provided by SL9} along with CS, \replaced{has been derived to be present}{was observed} up to the 2.0 mbar pressure region. The discrepancy of CS vertical distribution with \ce{H2O} and the decrease in its total mass may be attributed to the chemical loss of CS \replaced{at}{in} Jupiter's stratosphere lower than the 0.2 mbar altitude region. }
\end{abstract}



%% and \citet commands to identify citations. The citations are

%% their objects with \objectname{} or \object{}. Each macro takes the
%% differs from what is to be printed in the paper. The text appearing 
%% Otherwise, LaTeX will issue an error. 

%\clearpage

\section{Background}
\edit1{Cometary impacts} is considered to have affected the stratospheric composition of all four gas giant planets. On Jupiter, the 1994 collision of the fragments of comet Shoemaker-Levy 9 (referred to as the SL9 event) caused the delivery of \deleted{collision-induced}volatile gases such as CO, HCN, \ce{H2O}, \ce{S2}, \ce{H2S}, OCS, \ce{CS2} and CS \edit1{in} stratosphere (e.g. \cite{Noll1995}, \cite{Lellouch1995}, \cite{Moreno2003}). Recent sub-millimeter observations have revealed that the upper stratospheric CO molar fraction on Saturn, Uranus and Neptune is larger than that at the lower \replaced{region}{levels} (\cite{Lellouch2005}, \cite{Hesman2007}, \cite{Fletcher2010}, \cite{Cavalie2009}, \cite{Cavalie2014}) that indicates the presence of a CO external supply. On Neptune, since the measured CO/\ce{H2O} and CO/\ce{CO2} flux ratios are larger than the cometary [CO]/\ce{[H2O]} and [CO]/\ce{[CO2]} values, the \replaced{CO is likely to have originated by through the previous cometary impact and subsequent thermo-chemical process}{CO was probably formed during a comet impact through \edit1{shock chemistry}} rather than a steady influx of Inter-planetary Dust Particle (IDP) and subsequent evaporation \cite{Zahnle1996}. Moreover, observed vertical distribution of the stratospheric CO on Saturn and Uranus match with the cometary impact model, not the IDP model (\cite{Cavalie2010}, \cite{Cavalie2014}). 

\edit1{For the understanding of the influence of cometary impact event on the Jovian stratospheric composition, the detailed investigation of the time variation of the mass of the post-SL9 species on Jupiter yield valuable information}. We have focused on the \replaced{sulfur chemistry because S-bearing species may provide a signature of the cometary impact event on the gas giant planets}{most abundant sulfur species(CS) and its chemistry, because S-bearing species may be diagnostic of cometary impacts in other giant planets as well}. After the SL9 event, some S-bearing species were newly discovered in Jupiter's stratosphere \replaced{with}{at} various wavelengths ranging from ultraviolet to sub-millimeter. Within a month after the SL9 event, \ce{S2} was observed to be a dominant sulfur reservoir among the species discovered, such as \ce{CS2}, OCS, CS and possibly \ce{H2S} (\cite{Noll1995}, \cite{Lellouch1995}). The abundances of observed S-bearing species have shown different time variations such that while the spectral features of \ce{S2}, \ce{CS2}, OCS and \ce{H2S} had faded within a month (\cite{Noll1995}), CS molecule became a major sulfur reservoir instead. Since the total mass of \ce{S2} observed in 1994 as 10$^{14}$ g was comparable with that of \deleted{the} CS in 1995, CS was expected to be produced mainly by the conversion of \ce{S2} (\cite{Moreno2003}). To explain the conversion of \ce{S2} into CS, \cite{Moses1996} developed a sulfur chemistry model such that \ce{S2} can produce CS through a reaction with \ce{C2H3} as follows: \ce{S2 + C2H3 + H -> 2CS + 2H2}. HCS and \ce{H2CS} were expected to work as the intermediates to produce CS by the following reactions: \ce{HCS + H -> CS + H2}, \ce{H2CS + h$\nu$ -> CS + H2}. 
Frequent millimeter and sub-millimeter observations of \deleted{collision-induced} CS, HCN and CO have \added{been} performed from 1994 to 1998 to monitor their abundance, horizontal and vertical distributions by using the Institut de Radioastronomie Millim\'{e}trique (IRAM) 30-m telescope and 15-m James Clerk Maxwell Telescope (JCMT). Since the total mass of CS measured in 1998 as 1.4$\pm$0.6$\times$10$^{13}$ g was within the uncertainty of the value measured in 1995, no time variation of CS abundance was found during the monitoring period. The vertical distribution of CS was modeled as a simple two-layer model with a \ce{p_{0}} \replaced{transient}{cut-off} pressure level. In 1995, the \ce{p_{0}} value was derived to be 0.15 mbar at a longitude of 44$^{\circ}$S where \replaced{is the collision area}{the collision sites were located}. Owing to the vertical \replaced{transportation}{diffusion} of gases, the \ce{p_{0}} pressure level moved downward \added{to} approximately 0.2 mbar and 0.3 mbar in 1996 and 1998, respectively. \cite{Moreno2003} expected the \ce{p_{0}} value in 2014 to be 1.0 mbar according to the vertical transportation model. \deleted{The isotopomer of CS, \ce{C^{34}S} was also detected by using the JCMT, and the \ce{^{32}S}/\ce{^{34}S} was reported to be 1.4 -- 4.7 times higher than that of the terrestrial value(\cite{Matthews2002}).}

\replaced{Adding to}{Besides} Jupiter, several attempts have been made to detect S-bearing species on Saturn and Neptune(\cite{Fletcher2012}, \cite{Iino2014a}). However, negative observation results of \ce{SO2}, \ce{H2S} and CS on Saturn, and additional SO, OCS and \ce{C3S} on Neptune \replaced{were}{have been} reported. The discrepancies of these results with that of Jupiter is interesting since the \deleted{previous} cometary \replaced{impact}{impacts} on both planets have been suggested \replaced{also}{as the source of external CO in Saturn and Neptune}. 

\deleted{Photochemistry plays an important role in the atmospheric change of the planetary stratosphere. New observation to determine the long-term evolution of CS on Jupiter could yield an important information on the stratospheric sulfur chemistry \replaced{of gas giants}{in giant planets} because the 20-years elapsed time after the SL9 event significantly longer than the lifetime of CS against photolysis as 50-100 days at Jupiter's orbit (Sanzovo1993).} \added{A new observation of CS in Jupiter, 19 years after the SL9 event, could yield important information on the long-term evolution of its abundance and hence on the sulfur chemistry in the Jovian atmosphere.} Although the conversion process of \ce{S2} into CS occurring just after the impact has been fully explained already(\cite{Moses1996}), the sulfur chemistry \replaced{relating}{related} to the destruction\deleted{and recycling} of \edit1{the} CS is unknown. 

In 2013, \deleted{to obtain the time variation of CS abundance and vertical distribution on Jupiter, }we conducted \added{a} new observation \replaced{by using the sub-millimeter band, which is}{at submillimeter wavelength, which are} typically sensitive to \replaced{warm stratospheric gases}{the stratospheric temperature} \edit1{and composition}. In this paper, detail of the observation, data analysis and observation results are given in section 2. \replaced{Possible}{The} sulfur chemistry in Jupiter's stratosphere is discussed in section 3.
%\clearpage
\section{Methods}
\subsection{Observation}
Our observation was performed by using the Atacama Sub-millimeter Telescope Experiment (ASTE) 10-m single-dish telescope (\cite{Ezawa2004}), operated by National Astronomical Observatory of Japan (NAOJ), located at an altitude of 4800 m on the Atacama highland, Chile. Observation was performed from 9:36 UT to 11:33 UT, October 1, 2013. The angular diameter of Jupiter was 37.5'' whereas the full width half maximum (FWHM) of the ASTE at the 350 GHz band is 22''. The observation was conducted during the same period as our Neptune observation (\cite{Iino2014a}). 

We used a sideband separating mixer receiver CATS345 (\cite{Inoue2008}) that can be tuned from 324 GHz to 372 GHz. For CS, only the $J$=7--6 rotational transition is observable in the frequency range of CATS345. Separation of V and H polarization is not available during \added{2013} observation season. For the back-end, a MAC XF type digital spectrometer (\cite{Sorai2000}) was used with 512 MHz bandwidth and 500 kHz spectral resolution \added{mode} . Focus was \replaced{set}{calibrated} before the observing \edit1{run}. Similar to the previous observations \replaced{toward}{of} Jupiter and Saturn (\cite{Moreno2003}, \replaced{Cavalie2009}{\cite{Cavalie2010}}), both equatorial edges of Jupiter were set as ON and OFF positions to achieve the high signal to noise ratio \added{and remove as much as possible the standing waves induced by the observation of a strong and extended continuum source like Jupiter} . The integration time for each ON and OFF position was 15 s. For the pointing accuracy correction, cross-scan pointing observation of Jupiter's continuum emission was performed before and during the observing time. \added{The scanned direction was for azimuth and elevation, and pointing error for both the direction was calculated and calibrated automatically.} The measured pointing \replaced{accuracy was typically 0-1''}{error during the observation was 1'' and 0'' for azimuth and elevation, respectively,} which are better than the 2''-3'' typical pointing error of the ASTE. %\added{\footnote{http://alma.mtk.nao.ac.jp/aste/research_e.html}}. 
Owing to the low elevation angle during the observation as 39.5$^{\circ}$-44.5$^{\circ}$, the measured system temperature value of 532-627 K was higher than the typical value of ASTE of 250 K.

Just after the observation, one of the standard sources of ASTE, IRC10216, was observed to \replaced{check the receiver setting}{calibrate the receiver}. The obtained 6.1 K CS intensity of IRC10216 \replaced{shows 10$\%$ consistency}{is consistent within 10$\%$} with the previous single-sideband calibrated intensity of Caltech Submillimeter Observatory(CSO)(\cite{Wang1994}) obtained as 5.7$\pm$0.4 K in the antenna temperature scale. 
%\clearpage
\subsection{Observation results}
\begin{figure}
\begin{center}
\includegraphics[width=14cm, bb=0 0 595 595]{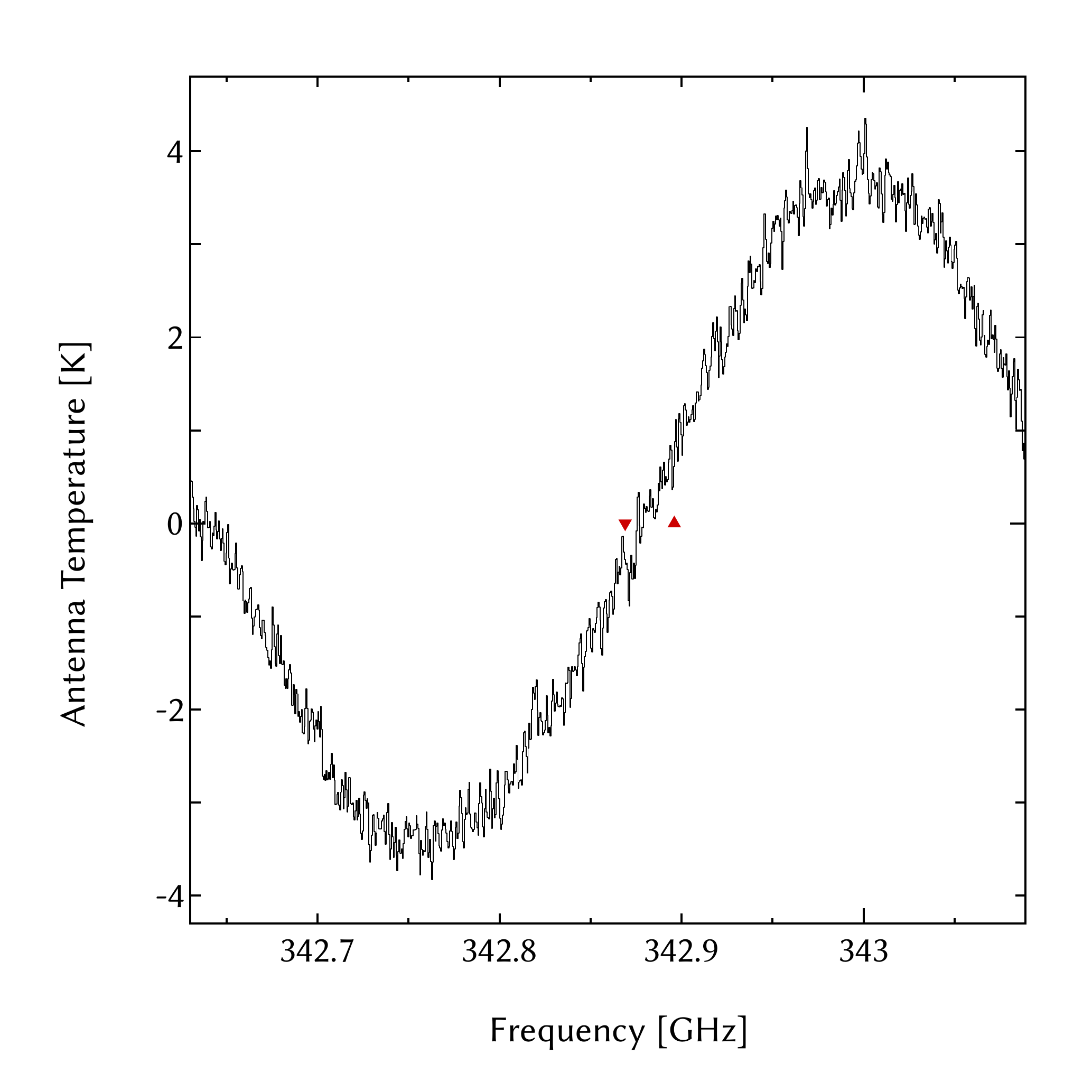}
\caption{\added{A raw spectrum for 15 seconds integration time. Each spectrum were averaged after the correction of the baseline structure by the polynomial function. \edit1{Red triangles indicate where the Doppler-shifted CS lines are expected. }}}
\label{fig:scan}
\end{center}
\end{figure}

\begin{figure}
\begin{center}
\includegraphics[width=18cm, bb=0 0 1191 595]{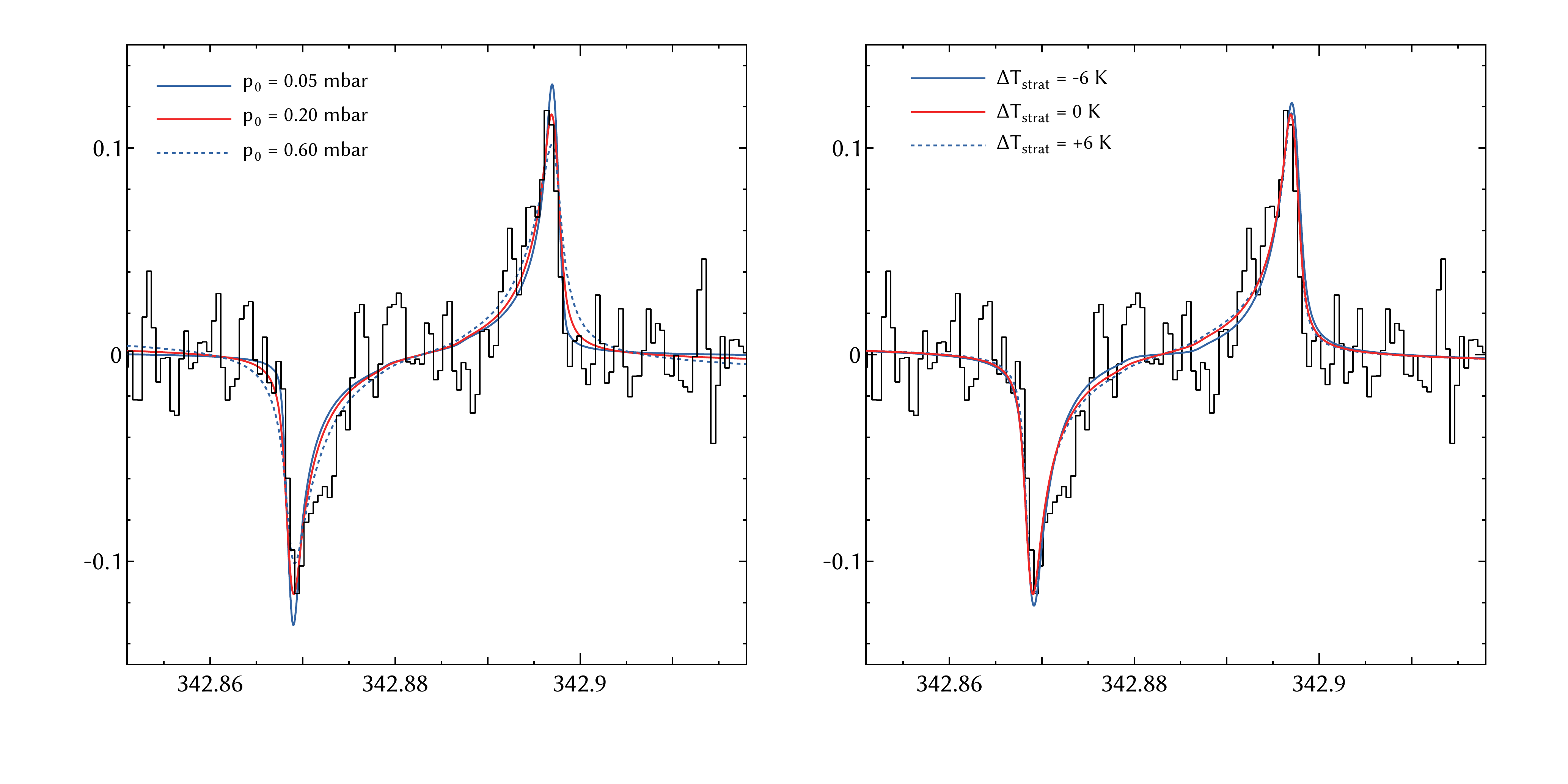}
\caption{\added{left panel:Observed (solid black line) and modeled spectra for \ce{p_{0}} = 0.05\added{(thin blue)}, 0.20\added{(thick red)} and 0.60 \added{(dotted blue)} mbar vertical distribution.} \deleted{\ce{p_{0}} = 1.00 mbar model(dashed black line) could not reproduce the observed spectra. } 
\added{right panel: Comparison of the best-fit spectra for +6 K (dotted blue) and -6 K (thin blue) stratospheric temperature ($\Delta$\ce{T_{strat}}). \ce{p_{0}} is fixed to 0.2 mbar. } }
\label{fig:spectra}
\end{center}
\end{figure}

\added{Figure \ref{fig:scan} shows one observation scan after the subtraction of OFF from ON integration.} The vertical and horizontal axes are expressed in the antenna temperature scale in units of Kelvin, and the rest frequency in GHz. To decrease the observed baseline structure, the polynomial fitting method was employed for each observation scans. Second, all of baseline-corrected scans equivalent to 47 min integration time were \replaced{integrated}{averaged}. The integrated spectrum is shown in Figure \ref{fig:spectra} in black lines. \replaced{The residual of the peak velocity is equivalent to the rotational velocity at Jupiter's equator.}{The lines seen in the ON and in the OFF spectra (in emission and absorption, respectively, in Figure \ref{fig:spectra}) are blue-shifted and redshifted (respectively) because of the planet rapid rotation.} The absorption structure is the \edit1{subtracted} emission of the OFF position. The obtained intensities for respective ON and OFF positions are $\sim$120 mK in the antenna temperature scale. The root-mean-square (RMS) noise level achieved here is 18.1 mK. Therefore, detection of the CS line was achieved with 6.6$\sigma$ confidence.\deleted{The observed intensity is $\sim$8 times lower than that of CS ($J$=7-6) observation result using JCMT reported in Matthews2002. Because the JCMT has smaller field of view of $\sim$14'' than that of ASTE of 22'', our obtained intensity should be weaker than JCMT owing to the beam dilution effect. }
%\clearpage
\subsection{\edit1{Radiative transfer} analysis of column density and vertical distribution of CS}
For the derivation of the CS abundance and vertical distribution, an atmospheric \edit1{radiative transfer} algorithm developed for terrestrial minor species (\cite{Ohyama2012}) was modified and applied to the spectrum. The rest frequency, line intensity and partition function of CS were extracted from the HITRAN 2012 spectroscopic database (\cite{Rothman2013}). The atmospheric structure employed here was the same as the previous observation (\cite{Matthews2002}). The relationship of the pressure broadening coefficient $\gamma$ with atmospheric pressure p and temperature T is described as follows: $\gamma$ = $\gamma$\ce{_{0}}$\times$(p/\ce{p_{0}})$\times$(\ce{T_{0}}/T)$^n$ cm$^{-1}$\edit1{$\cdot$bar$^{-1}$}, where \ce{p_{0}}, \ce{T_{0}} and n are 1 bar, 300 K and 0.75, respectively, and $\gamma$\ce{_0} is pressure broadening coefficient of CS at \ce{p_{0}} and \ce{T_{0}} as 0.125 cm$^{-1}$\edit1{$\cdot$bar$^{-1}$}. $\gamma$\ce{_0} value used here was that of used in the previous study of \cite{Moreno2003} (provided by R. Moreno, private communication). For the continuum opacity, the collision induced absorption (CIA) of \ce{H2}-\ce{H2} and \ce{H2}-He pairs were employed for each atmospheric layers. The values of CIA were obtained from the HITRAN catalogue. For the main beam efficiency value, we assumed the typical value of ASTE as 0.6. The spherical geometry was modeled to take into account the limb effect. \replaced{Radiative transfer calculation was performed for each grid that has 100 km spatial resolution}{1-D radiative transfer calculations have been performed on each gridpoints of a grid covering the planet, in which the spacing between gridpoints was set to 100 km (i.e. $\sim$1/720 equatorial radius)}. \added{The calculated spectra were averaged assuming the symmetric 2-D Gaussian beam pattern.} \added{The calculated opacities of the line center for the equatorial limb and disk center are 0.67 and 0.02, respectively. } Various \added{two-layer} vertical distribution models in which CS is present uniformly above the \ce{p_{0}} = 0.01-1.0 mbar pressure level were attempted to reproduce the obtained spectrum. Only the CS molar fraction was \edit1{varied} as the free parameters\added{, and the CS column density was derived from the profile.} 
%\clearpage
\subsection{Results of \edit1{radiative transfer}}

\deleted{The synthesized spectra for the modeled 0.05-0.60 mbar \ce{p_{0}} levels are represented in Figure \ref{fig:spectra} as red lines. } \deleted{The intensity of the \ce{p_{0}} = 1.0 mbar model was too low to reproduce the peak region of the observed spectrum.} \deleted{$\Delta \chi^2$ \deleted{method} was\replaced{employed}{calculated} to estimate the probable }
\deleted{Figure 2 shows the $\Delta \chi^2$ values for the \replaced{modeled each}{each modeled} \ce{p_{0}} \replaced{levels}{level}. } \edit1{From the analysis of the line width, we find that the most probable \ce{p_{0}} value}   \replaced{was}{is} 0.2 mbar, which \replaced{are}{is} consistent with \replaced{those}{the one} measured in 1998 as \added{0.3$^{+0.2}_{-0.1}$ mbar} (\cite{Moreno2003}). \deleted{The \ce{p_{0}} level corresponding to the }1$\sigma$ error of \ce{p_{0}} \replaced{is ranging}{range goes} from 0.05 to 0.6 mbar region. \deleted{Therefore, CS \deleted{molecule} is derived to be present above 0.6 mbar pressure level at least.} \edit1{Corresponding} \replaced{abundance}{mole fraction} and column density of \ce{p_{0}}=0.2 mbar model were 1.1$\times$10$^{-9}$ and 3.48$\times$10$^{13}$ molecules cm$^{-2}$, respectively. \added{The total mass of CS is 1.6$\times$10$^{12}$ g assuming \edit1{a} horizontally uniform CS distribution, and surface area with 71492 km equatorial and 61492 km polar radius (\cite{Archinal2011})}. 

\begin{figure}
\begin{center}
\includegraphics[width=14cm, bb=0 0 595 595]{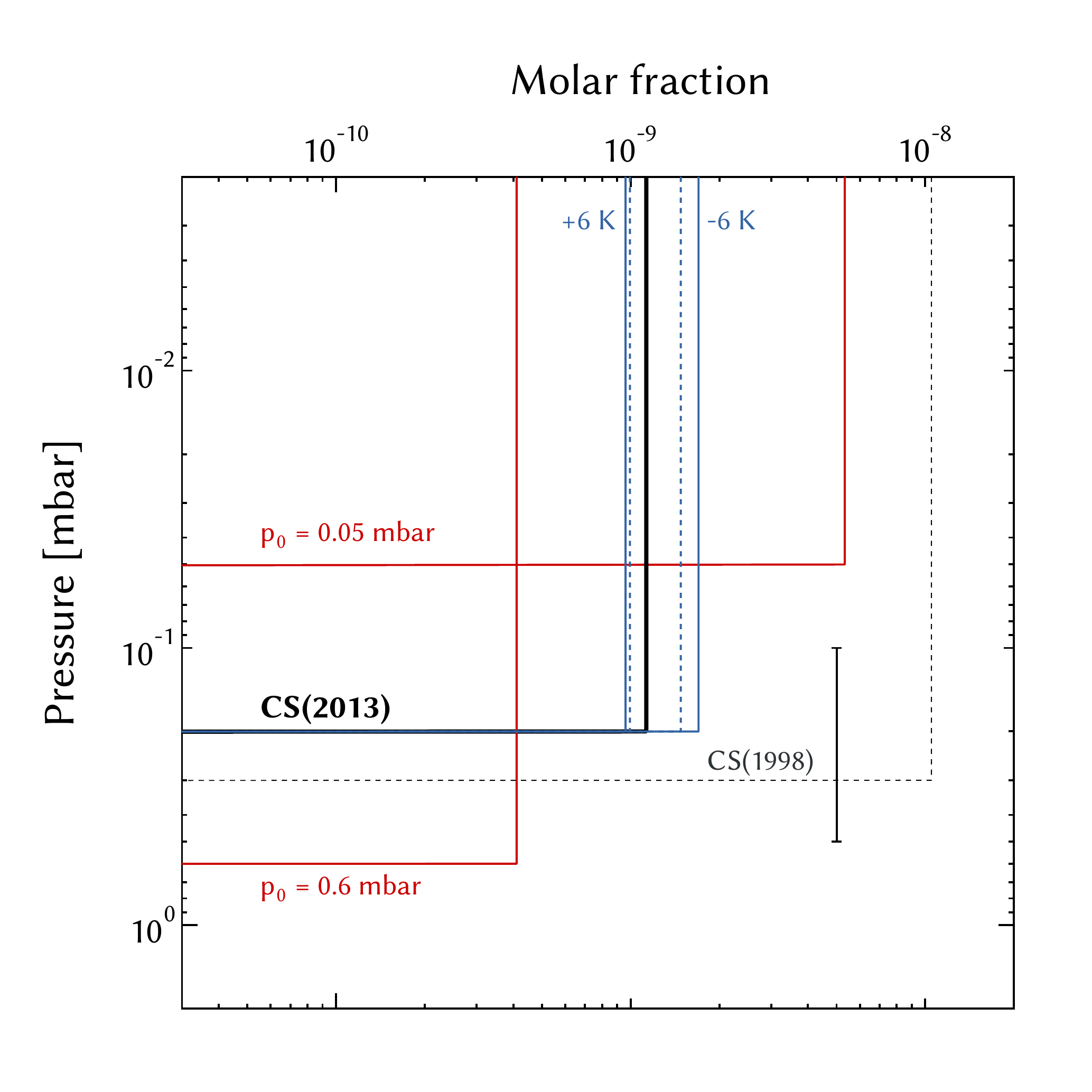}
\caption{\added{Thick black line shows the \edit1{best fit} CS vertical distribution model as \ce{p_{0}} = 0.20 mbar and 1.1$\times$10$^{-9}$ molar fraction. The \ce{p_{0}} value measured in 1998 \edit1{is} consistent with this work. Respective blue solid and dashed lines are for the $\pm$6 and $\pm$3 K stratospheric temperature models. CS profile measured in 1998 is plotted in thin black dashed line with its errors \edit1{on} \ce{p_{0}}.}
\deleted{Red lines represent the vertical distribution of CS for \ce{p_{0}} = 0.05(dashed), 0.20(solid) and 0.60(dashed) models. The results of CS obtained in 1998 (Matthews2002, Moreno2003) and those of \ce{H2O} in 2009/2010 (Cavalie2013) are shown by black dashed lines. Decrease of CS abundance was found in this study. Since CS is present in the upper region than \ce{H2O}, CS is likely to be ost selectively above 2.0 mbar region. The vertical distribution of CS expected by Moreno2003 in 2014 was above 1 mbar pressure, which is close to the observed distribution of \ce{H2O} in 2009/2010.}}
\label{fig:vertical_distribution}
\end{center}
\end{figure}

\deleted{Vertical distribution for the modeled \ce{p_{0}} levels are plotted in Figure \ref{fig:vertical_distribution} along with those of CS and \ce{H2O} measured in 1998 and 2009/2010, respectively (Moreno2003, Cavalie2013)} \added{We have tested the sensitivity of the column density result \edit1{against} the stratospheric temperature. Column densities were calculated for $\pm$6 and $\pm$3 K stratospheric temperature models 
which \edit1{includes} the previously obtained temparature variation from 160 to 166 K (\cite{Moreno2003}, \cite{Matthews2002}). Obtained errors were 3.5$^{+1.4}_{-0.8}$$\times$10$^{13}$ molecules cm$^{-2}$ and 3.5$^{+0.6}_{-0.6}$$\times$10$^{13}$ molecules cm$^{-2}$ for  $\pm$6 and $\pm$3 K models, respectively. These values are equivalent to $^{+41}_{-22}$ $\%$ and $^{+18}_{-17}$ $\%$ uncertainties.} CS vertical profile for the \ce{p_{0}} = 0.05, 0.20 and 0.60 mbar models, $\pm$6 and $\pm$3 K stratospheric temperature models are plotted in Figure \ref{fig:vertical_distribution} along with that of CS measured in 1998 (\cite{Moreno2003}). 
%\clearpage

\section{Discussion}

\deleted{Presence of the recycling system of CS in Jupiter's stratosphere}

In May 1995, 10 months after the SL9 event, the CS column density measured at 44$^{\circ}$S at the latitude of comet collision was reported as 0.6-1.4$\times$10$^{15}$ cm$^{-2}$ \added{\edit1{corresponding to a total mass of 2.0$\pm$1.0$\times$10$^{13}$ g} (\cite{Moreno2003})}. \added{During the long-term monitoring of CS from 1995 to 1998, its total mass did not show \edit1{a} significant change, and the \edit1{best fit} value throughout the period was reported as 1.8$\pm$0.4$\times$10$^{13}$ g.} \deleted{After the horizontal dispersion of CS \added{in xxx}, the \replaced{global}{disk-averaged} column density was expected to be 5$\times$10$^{14}$ cm$^{-2}$ in the same paper \added{, equivalent to 1.8$\times$10$^{13}$ g total mass}. } Thus, our new observation shows that the CS \replaced{column density}{total mass} \added{measured in 2013} decreased \edit1{by} up to $\sim$90$\%$ \edit1{with respect to} \replaced{the previous estimation}{that of in 1998}. Due to the low S/N, \edit1{we have not attempted to constrain the temporal variation of the vertical distribution of CS and thus eddy mixing in Jupiter's stratosphere.}

\edit1{Explaining the decrease} of the CS column density is important for the understanding of the sulfur chemistry in the stratosphere of gas giants. \deleted{To \replaced{constraint}{constrain} the removal process of CS, a comparison of the derived vertical distribution result with the case of other trace species produced by the SL9 event in water vapor is helpful. Water vapor was first discovered by the Infrared Space Observatory (ISO) satellite in 1997 with 2.0$\pm$0.5 $\times$10$^{14}$ cm$^{-2}$ global column density (Lellouch1997), and \ce{p_{0}} level was estimated to be 0.5$\pm$0.2 mbar. Lellouch1997 concluded that the origin of this water vapor is a sporadic supply process such as the SL9 event rather than a continuous supply such as an influx of IDP. In 2009 and 2010, Cavalie2013 derived both the vertical and horizontal distribution of water vapor in Jupiter's stratosphere with the $Herschel$ satellite. The \ce{p_{0}} level determined by the fine spectral resolution achieved by the HIFI\deleted{ and PACS radiometer }was 2 mbar with a 1.7$\times$10$^{-8}$ mixing ratio \replaced{of}{in} the upper layer. 
}
Condensation or polymerization processes may contribute to the removal of CS from the gas phase to produce solid or liquid phase, or polymerized CS expressed as (CS)\ce{_n}. \edit1{Phase change or polymerization of CS could occur in the stratosphere as temperature decreases and number density increases with decreasing altitude.}  However, the precise evaluation of the phase change is unavailable due to the absence of phase change parameters in the literature. Polymerization is a characteristic phenomenon of S-bearing species. In particular, polymerization of CS on the aerosol surface, which would occur and remove the gas phase CS. 
\deleted{The other scenario is such that other sulfur reservoir may be more abundant gas-phase species other than CS below the 0.6 mbar region. Since the ultraviolet shielding works more efficiently in the lower stratosphere, photolysis-induced hydrocarbon radicals are less abundant, and the aforementioned CS recycling processes do not work efficiently. Gladstone1996 expected that the abundances of dominant \ce{C1} and \ce{C2} hydrocarbon radicals such as \ce{CH3}, \ce{C2H3} and \ce{C2H5} show nearly a proportional decrease in relative to the atmospheric pressure. In such a case, neutral-neutral reactions may work more efficiently than the photolysis and/or reactions with radicals. For example, reaction of the S-atom with neutral molecule such as \ce{H2} and \ce{CH4} that produce HS radical may lead to the production of other sulfur molecules such as \ce{S2}, \ce{CS2} and \ce{H2S}. The other reaction \ce{S + CO -> OCS} may be important because CO is also a co-product of the SL9 event along with CS, HCN and water vapor. }
\added{The other scenario is that other S-bearing species works as the sulfur reservoir in the deep stratosphere. However, in 1995 and 1996, while some S-bearing species such as \ce{H2CS}, NS, SO, \ce{H2S} and HNCS were searched already in millimeter waveband, none of them were detected (\cite{Moreno2003}).}

From the photochemical perspective, the discovery of CS even in 19 years after the production is interesting because CS is considered to be a chemically unstable molecule due to its short lifetime against the photolysis \added{and high reactivity}. \deleted{In this section, we discuss both the recycling and dissipation process of CS in the stratosphere of Jupiter.} \deleted{Preservation of CS in Jupiter's stratosphere} Photolysis (\ce{CS + h$\nu$ -> C + S}) is the rapid destruction process of CS in the gas phase. \deleted{Since the lifetime of CS against photolysis in Jupiter's orbit of 5 AU was estimated to be as short as 50-100 days (Sanzovo1993), considering only the photolysis, CS should has been dissipated completely in 19 years.} Therefore, the discovery of a CS molecule on Jupiter suggests the presence of \added{a strong ultraviolet shielding to prevent the CS photolysis or }a recycling process for CS. \added{Photodissociation of CS occurs at wavelength $\lambda$ = 1362, 1401 and 1542 \r{A} (\cite{Sanzovo1993}). Since these absorption wavelength is covered with the photodissociation band of major hydrocarbons such as \ce{CH4}, \ce{C2H4} and \ce{C2H6} (\cite{Gladstone1996}), CS photodissociation may be prevented efficiently. Even if the photodissociation occurs, dissociated S-atom could react with hydrocarbon radical rapidly as follows: \ce{S + CH3 -> H2CS + H}, \ce{S + CH2 -> HCS + H}. Produced \ce{H2CS} and HCS probably photodissociate to produce CS. }

\deleted{
Due to the rapid photolysis of CS, the recycling process of CS is initiated from dissociated S-atom. Reaction with \ce{H2} is an hydrogen abstraction process that produces HS radical and H-atom described as follows: \ce{S + H2 -> HS + H}. Although \ce{H2} is the dominant molecule in Jupiter's atmosphere, this reaction is negligible due to the significant reaction barrier in the low temperature environment (Leen1988). Reaction with radicals such as \ce{CH3}, \ce{CH2}, CH, \ce{C2H}, \ce{C2H3} \ce{C2H5} and OH should be considered consequently. Gladstone1996 investigated the hydrocarbon photochemistry in detail including the vertical diffusion and ultraviolet shielding. Among the hydrocarbon radicals, \ce{CH3} is expected to be the most abundant one. Detection of \ce{CH3} has been achieved by the $Cassini$ spacecraft during its 2000-2001 flyby in Jupiter's auroral infrared hot spots along with hydrocarbons such as \ce{CH4}, \ce{C2H2}, \ce{C6H6} and \ce{C2H6}(Kunde2004a). No other hydrocarbon radicals have been detected yet in the stratosphere thus far, and the predicted \replaced{abundance}{abundances} of \replaced{\ce{^{3}CH2}}{\ce{CH2}}, \ce{C2H3} and \ce{C2H5} radicals reported by Gladstone1996 was small as $\sim$1$\%$ of \ce{CH3}. The main reactions of the S-atom with hydrocarbon radicals incorporated in the chemical reaction databases of the University of Manchester Institute for Science and Technology (UMIST) and KInetic Database for Astrochemistry (KIDA) databases\added{(Wakelam2012)} are as follows:
}

\deleted{S + CH3 -> H2CS + H ,
S + CH2 -> HCS + H,
S + CH2 -> CS + H2.
}

\deleted{
The production of \ce{H2CS} might play an important role in the CS production because its photolysis produces CS as follows: \ce{H2CS + h$\nu$ -> CS + H2} \replaced{as mentioned in Section 1}{Moses1996}. The HCS radical may also contribute to the CS production via the following hydrogen abstraction, reaction with the H-atom and the following photolysis: \ce{HCS + H2 -> H2CS + H}, \ce{HCS + H -> CS + H2}, \ce{HCS + h$\nu$ -> CS + H}. The reactions with more complicated hydrocarbon radicals such as \ce{C2H3} and \ce{C2H5}, which might produce \ce{H2C2S} and \ce{H3C2S}, respectively, are not incorporated in the databases. The reaction of the S-atom with the OH radical produces SO as follows: \ce{S + OH -> SO + H}. However, although the high reactivity, the production of SO is negligible owing to the low mixing ratio of the mother species \ce{H2O} as 2$\times$10$^{-8}$ above 2 mbar pressure level (Cavalie2013). 
The only reaction of CS with a hydrocarbon radical incorporated in the databases is \ce{CS + C2H -> C3S + H}, whereas \ce{C2H} is not abundant among other radicals. The reaction of CS with the H-atom needed to produce HCS is an exothermic process that leads to the recycling of CS by the reactions mentioned above. 
In summary, the CS production process via \ce{H2CS} and HCS might recycle CS efficiently in Jupiter's hydrocarbon-rich stratosphere. 
%\clearpage
}

\deleted{Removal processes of CS}

\deleted{
It should be pointed out that the \ce{p_{0}} level of \ce{H2O} measured in 2009 and 2010 was deeper than that of the CS measured in several years after the last \ce{H2O} observation. This discrepancy in the vertical distribution of the two species may be helpful for the constraint the removal process of CS. Two types of diffusion processes are present in the planetary atmosphere: molecular diffusion and eddy diffusion. Above the homopause, the molecular diffusion could explain the differences in \ce{p_{0}} level between the two molecules . Below the homopause, however, the eddy diffusion plays a more important role than molecular diffusion, and the two molecules should show the same vertical distribution because the molecular flux induced by the eddy diffusion does not depend on the sort of molecule. Various attempts have been made to reproduce the observed hydrocarbon vertical distribution by using the photochemical and vertical transportation model including both diffusion processes. Such attempts have shown that the \ce{CH4} homopause is expected to be present at a pressure region of 10$^{-4}$ to 10$^{-3}$ mbar which is significantly higher in altitude than the \ce{p_{0}} level of CS and \ce{H2O} (Moses2005, Gladstone1996, Young et al. 2005). Assuming that both CS and \ce{H2O} were produced at the same time and considering only eddy diffusion, both molecules are expected to have the same \ce{p_{0}} level. The new discovery of the different \ce{p_{0}} level leads to a new scenario such that gaseous CS is being removed particularly in the p $>$ 0.6 mbar region. 
}

\deleted{Future}
\deleted{Further observation of other sulfur species is important for evaluating the scenarios given in this study, and a new search of \ce{H2CS} will lead to the evaluation of the possible CS recycling processes above} In the case of Neptune, a previous large cometary impact is \deleted{strongly}suggested by the larger \edit1{external} flux of CO than that of water vapor (\cite{Lellouch2005}, \cite{Hesman2007}). \edit1{The absence of CS (and other S-bearing molecules), as found with ASTE (\cite{Iino2014a}), could contradict the comet impact hypothesis. However, a detailed sulfur chemistry model should be developed for the further assessment whether the absence of CS is compatible with a comet impact scenario on Neptune (and also on Saturn). Such a model would be helpful in understanding the temporal evolution of the CS mass in Jupiter's stratosphere.} \added{The new development of the detailed sulfur chemistry model includes the precise estimation of ultraviolet shielding and other photochemical processes is helpful.} Once the sulfur chemistry on Jupiter is revealed, the result will be \replaced{applied}{a good reference} not only to Neptune but also to Saturn\deleted{and Uranus} where the previous cometary impacts are suggested. 

\acknowledgments
\added{Authors thank Thibault Cavali{\'{e}} for his fruitful referee comment that improved overall the content and scientific clarity. } Raphael Moreno of Observatoire de Paris is appreciated for offering advice and providing the pressure-broadening coefficient of CS, which was of significant help for our study. The observation was supported by Chihaya Kato of ISEE and Chihomi Hara of University of Tokyo. Proposal writing was supported and advised by Akira Mizuno, Takehiro Hidemori and Satonori Nozawa of ISEE. TI is also grateful for the continuous encouragement given by Takao Kokugan, Toshiyuki Watanabe, Yongbo Li and Tetsuya Mizutani of TUAT. The ASTE telescope is operated by NAOJ. The authors are grateful to the ASTE team for providing observation support and data handling. This work was financially supported by the department of organic and polymer materials chemistry of TUAT, and the leading graduate school program of Nagoya University.

\clearpage

\clearpage

\clearpage

%\begin{figure}
%\begin{center}
%\label{fig:delta_chi_square}
%\includegraphics[width=14cm, bb=0 0 567 567]{delta-chi-square.pdf
%\caption{$\Delta \chi^2$ values of the synthesized spectra for the modeled \ce{p_{0}} levels ranging from from 0.05-1.0 mbar. The most probable values are located at the 0.2 mbar region. The dotted horizontal line represents $\Delta \chi^2$=1, which is equivalent to 1$\sigma$ error of \ce{p_{0}}. 
%}
%\end{center}
%\end{figure}\input{iino2016b_rev1.bbl}

\clearpage

\begin{figure}
\begin{center}
\includegraphics[width=14cm, bb=0 0 595 595]{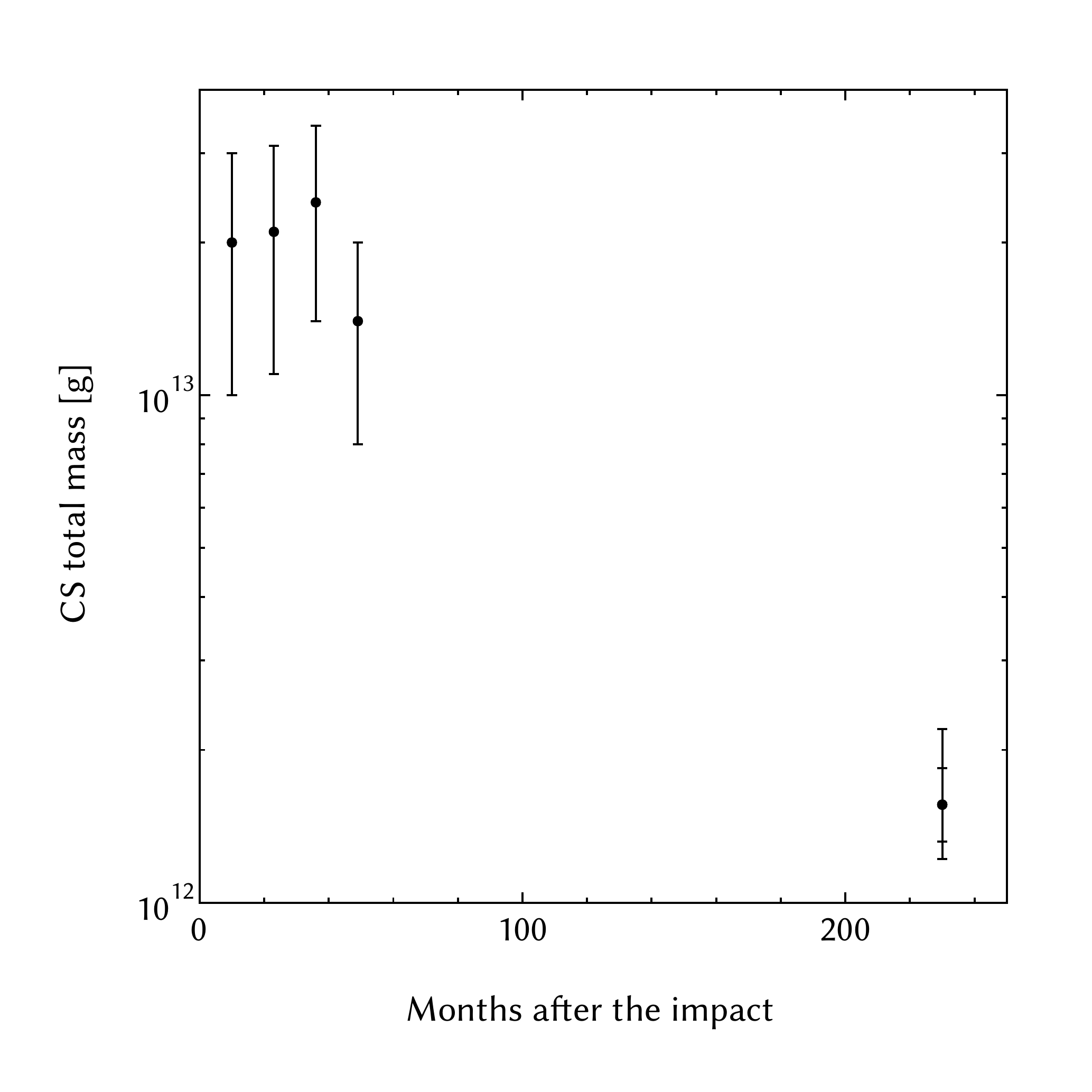}
\caption{\added{The time variation of CS total mass from 1995 to 2013 plotted in logarithm with respect to the months after the SL9 impact. Two error bars in this work corresponds to $\pm$3 and $\pm$6 K stratospheric temperature. CS total mass measured in 2013 shows $\sim$90$\%$ decrease in comparison with from 10 to 49 months period.}}
\label{fig:time_variation}
\end{center}
\end{figure}

%% The following command ends your manuscript. LaTeX will ignore any text

\clearpage

\end{document}